\documentclass[12pt]{iopart}

\newcommand{\jpsi}{\rm J/$\psi$}
\newcommand{\psip}{$\psi^\prime$}

\usepackage{graphicx}

\begin{document}

\title[J/$\psi$ production in \mbox{In-In} and \mbox{p-A} collisions]
{J/$\psi$ production in \mbox{In-In} and \mbox{p-A} collisions}

\author{E Scomparin for the NA60 Collaboration\footnote{For the full list of 
NA60 authors and acknowledgements, see Appendix `Collaborations' of this volume}\\
}
\address{INFN and Universit\`a di Torino, Via P. Giuria 1, I-10125 Torino
(Italy)}
\ead{scomparin@to.infn.it}
\begin{abstract}

The NA60 experiment studies dimuon production in \mbox{In-In} and \mbox{p-A} 
collisions at the CERN SPS. We report recent results on \jpsi\ production, 
measured through its muon pair decay. As a function of centrality, we show that 
in \mbox{In-In} the \jpsi\ yield is suppressed beyond expectations from nuclear 
absorption. We present also for the first time results on \jpsi\ production in 
\mbox{p-A} collisions at 158 GeV, the same energy of the nucleus-nucleus data. 
For both \mbox{p-A} and \mbox{In-In} we show preliminary results on \psip\ 
suppression. Finally, we have studied the kinematical distributions of the 
\jpsi\ produced in \mbox{In-In} collisions. 
We present results on transverse momentum and rapidity, as well as on the 
angular distribution of the \jpsi\ decay products.

\end{abstract}


\section{Introduction and data analysis}
The production of the charmonium states in ultrarelativistic heavy-ion collisions has been 
recognized as a key observable for the occurrence of deconfinement since the very 
beginning of the studies in this field~\cite{Sat86}. A suppression
of the charmonium yield beyond expectations from a pure nuclear absorption scenario 
has been observed both at SPS and RHIC energies by the NA38/NA50~\cite{Ale05} and 
PHENIX~\cite{Ada06} experiments, respectively. 
However, the interpretation of the experimental observations is still 
under debate, with explanations including, among others, the production of a fully 
thermalized QGP~\cite{Gra04}, of a percolating partonic condensate~\cite{Dig04}, or of a 
very dense hadron gas~\cite{Cap05,Mai05}. 
Therefore, accurate experimental investigations, carried out with several
projectile-target systems, are mandatory in order to confirm or exclude the various 
proposed scenarios. Clearly, in order to claim a suppression signal beyond nuclear 
absorption, accurate \mbox{p-A} reference data, taken in the same energy/kinematical 
conditions as those for the nucleus-nucleus data, are also necessary.  
In parallel to these studies, the investigation of the kinematical distributions of
charmonia provides interesting and complementary insights to the physics picture, giving
information on production mechanisms and initial state effects.

All of these aspects of charmonia production are presently being studied, at the SPS, 
by the NA60 experiment. The apparatus consists of a muon spectrometer, inherited
from NA50,
which is also used for triggering on the production of a muon pair. A vertex 
spectrometer, made of 16 Si pixel planes, tracks the charged particles produced in the
angular acceptance of the muon spectrometer (roughly 0$<y_{\rm CM}<$1 for a 158 GeV/c incident 
beam), giving precise 
information on the position of the interaction vertex ($\sim 20
\mu$m in the transverse direction, $\sim 200 \mu$m in the longitudinal one). Furthermore,
by matching muon tracks detected in the muon spectrometer with the corresponding tracks
in the vertex spectrometer, it is possible to i) significantly improve the invariant mass
resolution (from $~\sim$70 to $\sim$20 MeV at the $\omega$, from $\sim$100 to $\sim$70 MeV at 
the \jpsi) and ii) determine, with a $\sim$40 $\mu$m resolution, the
offset of the production point of the muons with respect to the interaction vertex. 
Finally, a Zero-Degree Calorimeter is used to provide a centrality selection through the
measurement of the energy $E_{\rm ZDC}$ carried away by spectator nucleons. The whole 
apparatus,  including details on the matching procedure, is 
described in~\cite{Ban05,Usa05,Sha06}. NA60 has studied \mbox{In-In} collisions
at 158 GeV/nucleon and \mbox{p-A} collisions at 158 and 400 GeV. The results presented
here refer to \mbox{In-In} and \mbox{p-A} at 158 GeV, while the analysis of the 400 GeV
data sample is still in progress. The event selection procedures, as well as details on 
the acceptance calculation and on the extraction of the physics signal, can be found 
in~\cite{Arn07}. After data reduction, we end up with about 3$\cdot$10$^4$ \jpsi\ events in 
\mbox{In-In} and a similar number in \mbox{p-A}.  
  
\section{\jpsi\ {\bf suppression in \mbox{In-In} collisions}}
\begin{figure}[h]
\centering
\resizebox{0.45\textwidth}{!}
{\includegraphics*{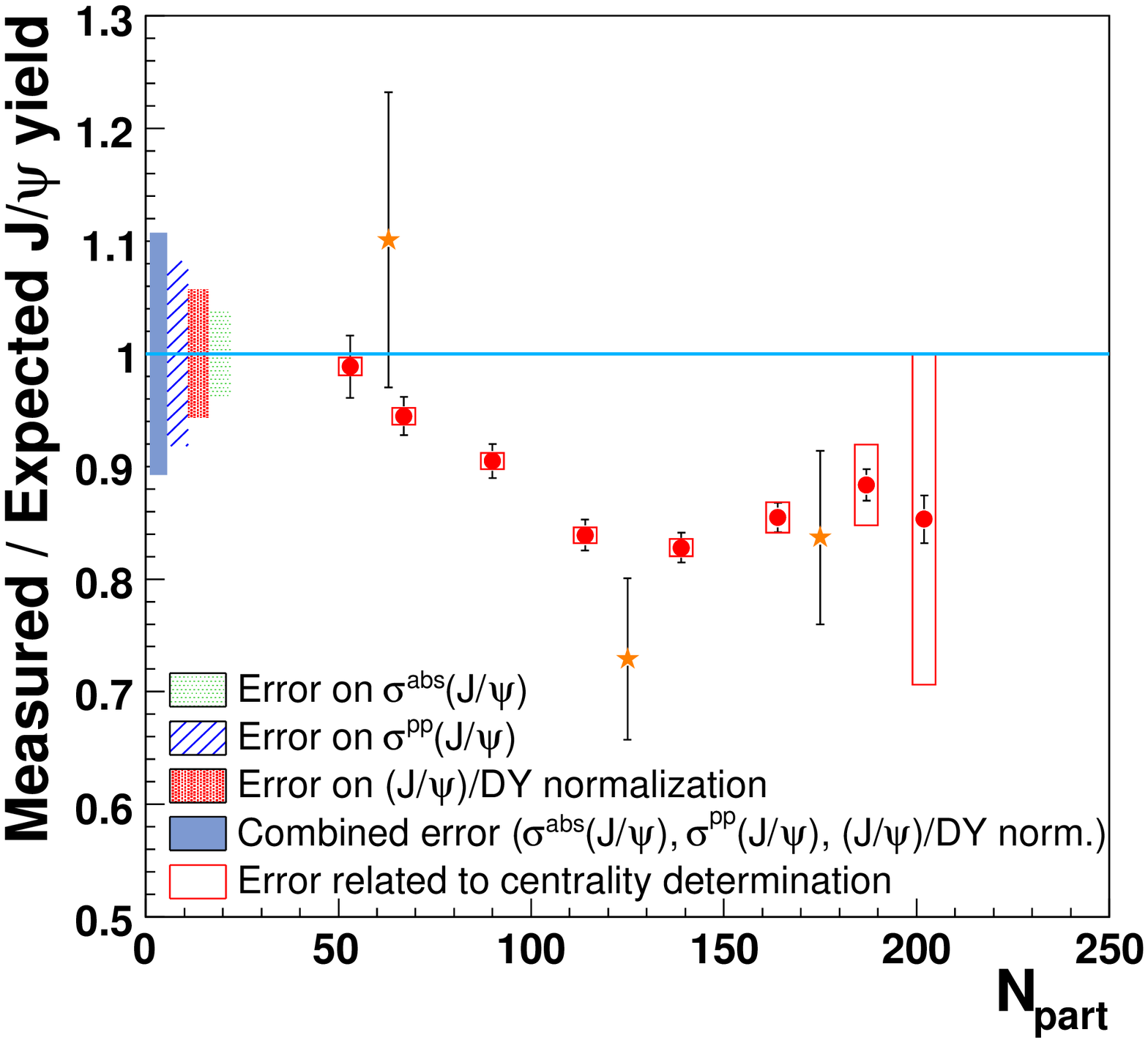}}
\resizebox{0.45\textwidth}{!}
{\includegraphics*{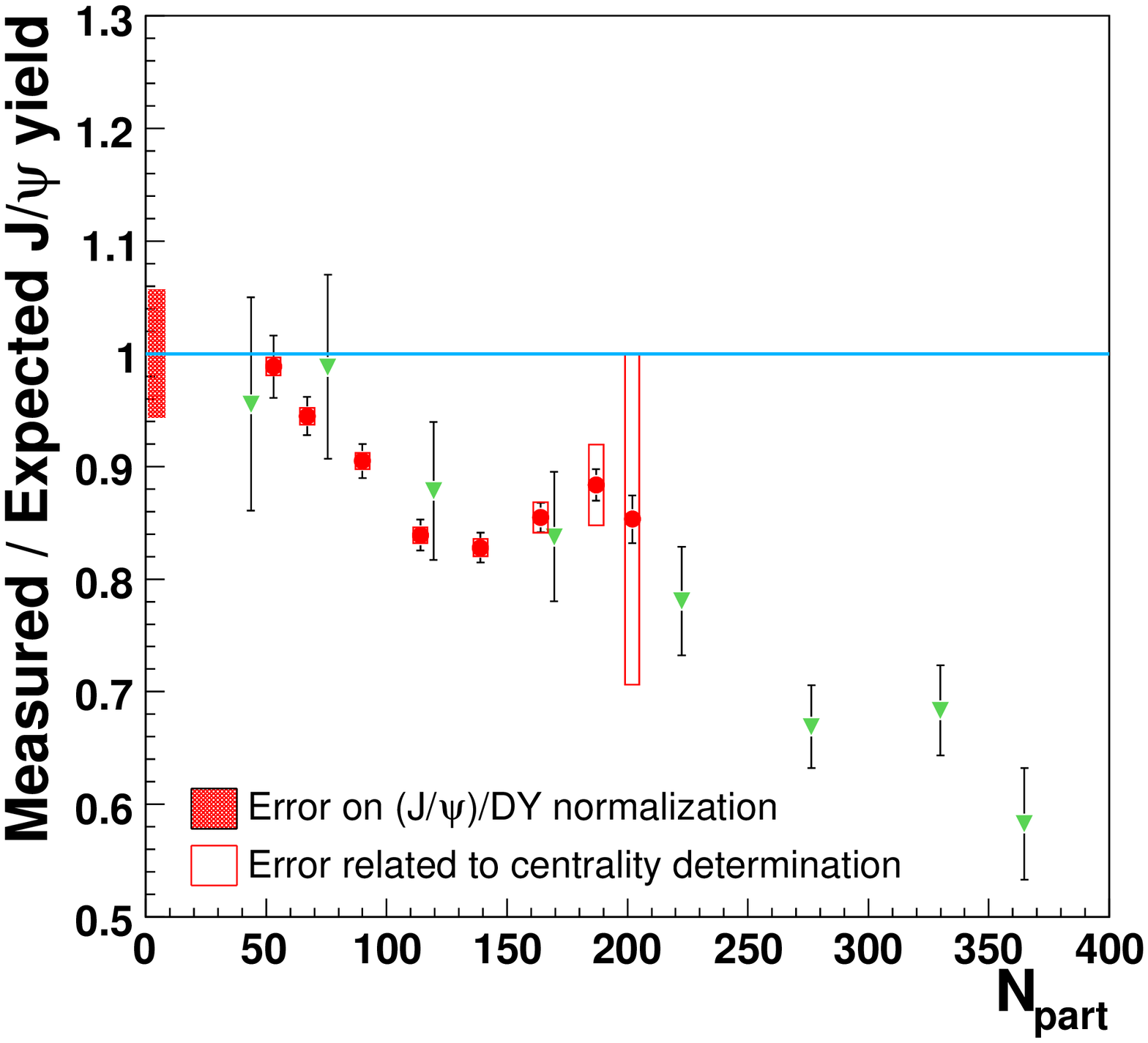}}
\caption{(left) Centrality dependence of the \jpsi\ suppression measured in \mbox{In-In} 
collisions. The stars correspond to the ratio between measured and expected 
$\sigma_{\rm J/\psi}/\sigma_{\rm DY}$, while the circles refer to the ratio between the 
measured \jpsi\ yield and nuclear absorption calculations. Systematic errors are also 
shown. (right) Comparison between the \mbox{In-In} (NA60, circles) and \mbox{Pb-Pb} 
(NA50, triangles) suppression patterns.}
\label{fig:1}
\end{figure}

The study of the \jpsi\ suppression in \mbox{In-In} has been carried out using two
different and complementary approaches. In the first, identical to the one adopted by 
NA38/NA50, the \jpsi\ yield has been normalized to the measured Drell-Yan events in the 
mass region $2.9<m_{\mu\mu}<4.5$ GeV/c$^2$. This quantity has the advantage of being free
from the systematic errors connected with efficiency and luminosity calculations, but
suffers from the low Drell-Yan statistics. In NA60, this analysis can be meaningfully
performed only with a very limited number of centrality bins. The ratios 
$\sigma_{\rm J/\psi}/\sigma_{\rm DY}$ are then compared with the expected values in case
of pure nuclear absorption. Such values have been obtained with the Glauber model, 
starting from the value $\sigma^{\rm abs}_{\rm J/\psi}$=4.18 $\pm$ 0.35 mb, measured by 
NA50 \cite{Ale05} in \mbox{p-A} collisions at 450 GeV. The result is plotted in
Fig.~\ref{fig:1}(left). Clearly, to increase the statistical significance, the
use of Drell-Yan should be avoided. This choice is the foundation of the second analysis
approach, where the measured ${\rm d}N_{{\rm J}/{\psi}}/{\rm d}E_{\rm ZDC}$ 
has been directly compared to a calculated reference spectrum corresponding to a pure nuclear 
absorption scenario. The shape of such a reference has again been calculated in the frame of 
the Glauber model, with $\sigma^{\rm abs}_{\rm J/\psi}$=4.18 mb.
We do not have yet for our \mbox{In-In} data an absolute determination of the 
cross section ${\rm d}\sigma_{{\rm J}/{\psi}}/{\rm d}E_{\rm ZDC}$. Therefore, 
we simply require the relative normalization between the measured and expected distributions to 
be equal, when integrated over centrality, to the same quantity obtained from the study of 
$\sigma_{\rm J/\psi}/\sigma_{\rm DY}$, i.e. 0.87$\pm$0.05. The result of this analysis is
also plotted in Fig.~\ref{fig:1}(left). Of course, the agreement between the results of the two analyses 
is significant only in terms of shape, since the normalization of the second analysis has been forced 
to be the same of the first one. Since with this approach the statistical errors are
negligible ($\sim$2\%, with the chosen centrality binning) a careful estimate of the
systematic errors is mandatory~\cite{Arn07}. It turns out that there is a $\sim$ 10\% error,
independent of centrality, essentially due to uncertainties in the Glauber model
parameters and in our knowledge of the inputs that enter in the nuclear absorption
calculation. On top of that, (small) uncertitudes on the link between $E_{\rm ZDC}$ and
the number of participant nucleons $N_{\rm part}$, due to the contribution of 
non-spectator energy to the measured signal, 
induce a non-negligible systematic error for very central events.  
Of course, most effects discussed here also affect the determination 
of $\sigma_{\rm J/\psi}/\sigma_{\rm DY}$, although their effect in absolute terms is in
this case much less important. The result plotted in Fig.~\ref{fig:1}(left) clearly indicates 
an anomalous suppression of the \jpsi\ yield for $N_{\rm part}\,>\,80$, with a saturation of
the effect for central \mbox{In-In} collisions. In Fig.~\ref{fig:1}(right) we compare the
suppression pattern obtained by NA60 with the NA50 results for \mbox{Pb-Pb}
collisions~\cite{Ale05}.
Within errors, the two behaviours look compatible, showing that $N_{\rm part}$ could be
 a good scaling variable for the onset of the anomalous suppression.  

\begin{figure}[h]
\centering
\resizebox{0.45\textwidth}{!}
{\includegraphics*{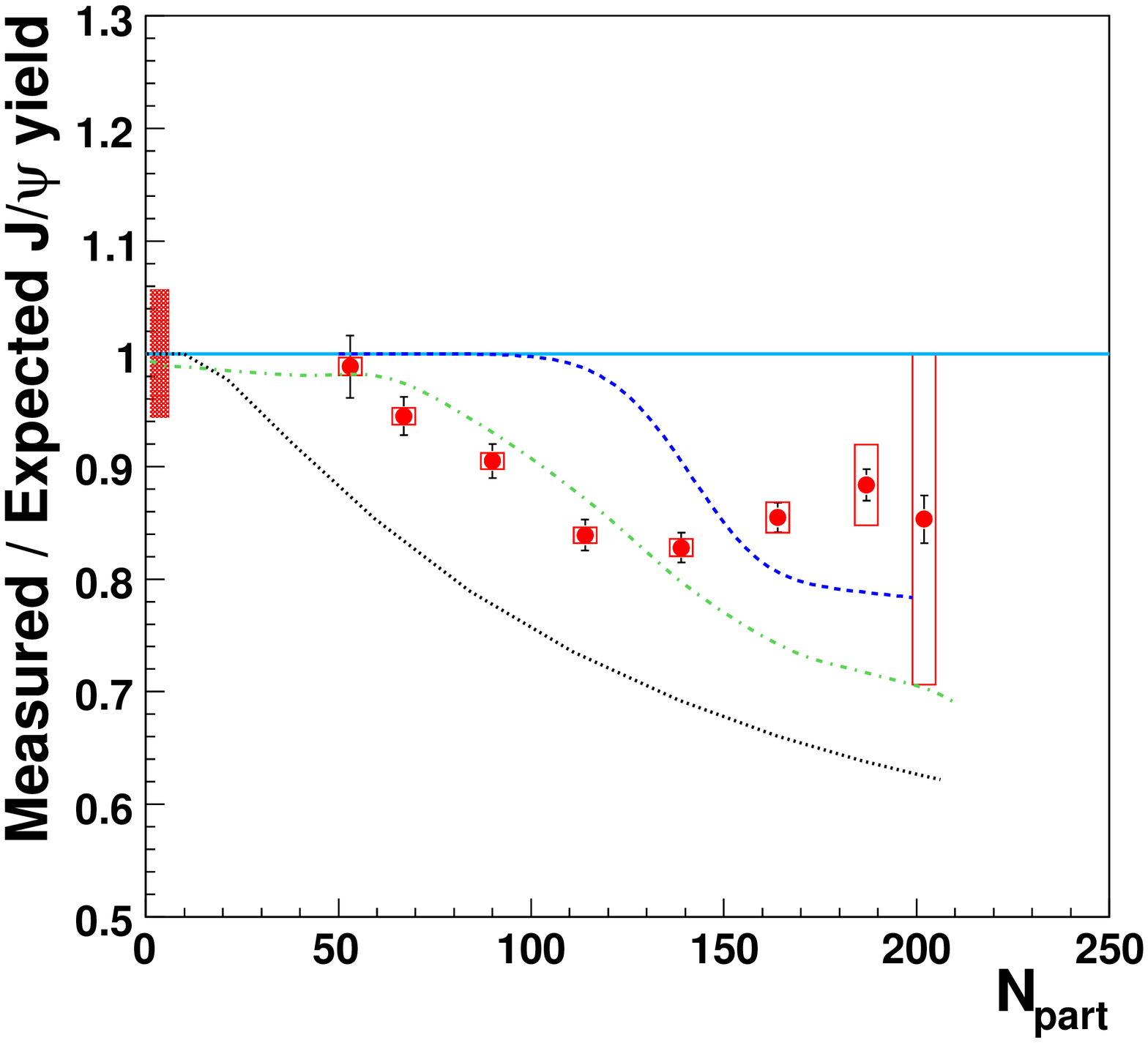}}
\resizebox{0.45\textwidth}{!}
{\includegraphics*{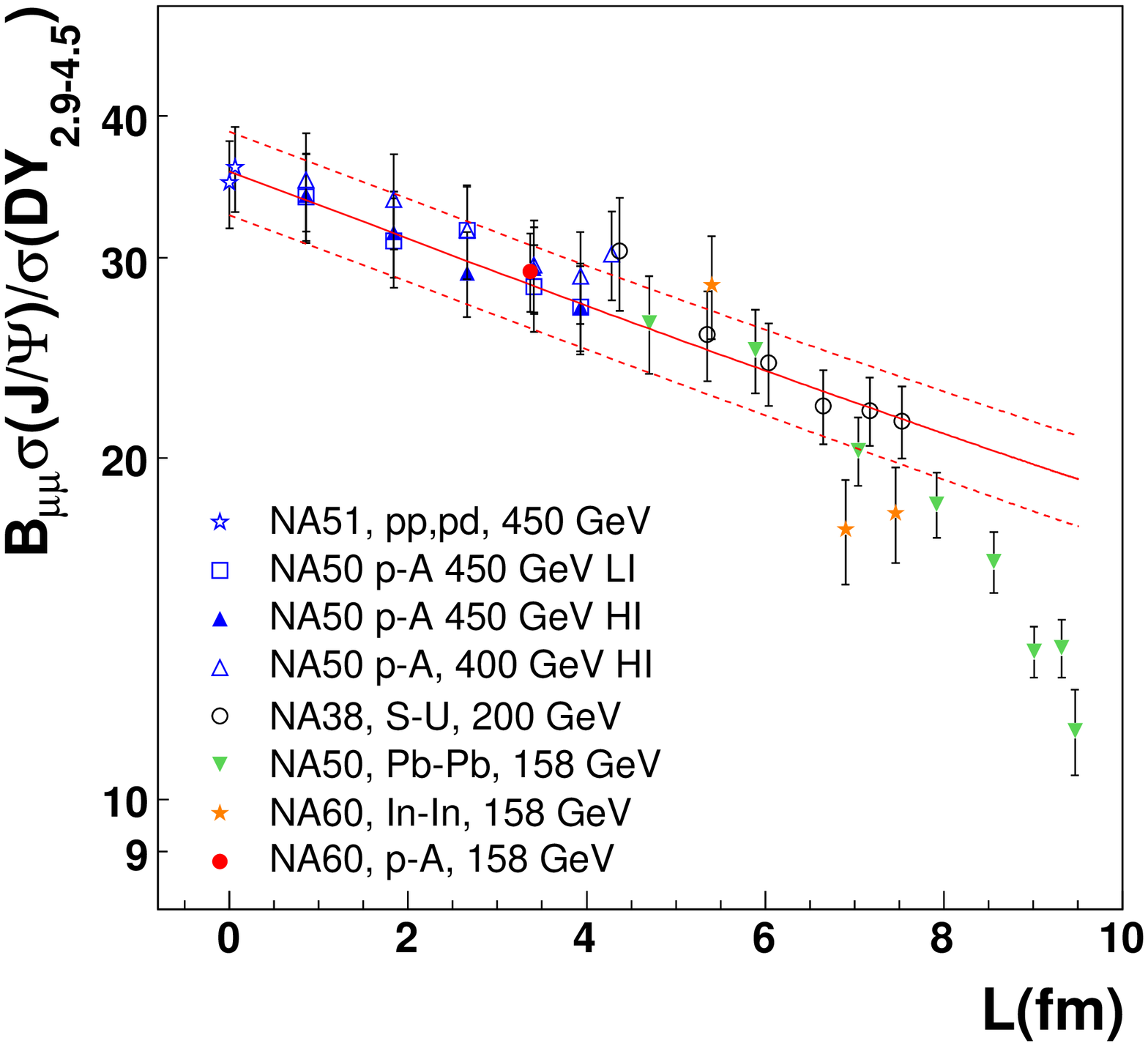}}
\caption{(left) Comparison between the \mbox{In-In} suppression pattern and the
theoretical predictions of Ref.~\cite{Cap05}(dotted line), \cite{Rap05}(dashed-dotted
line), \cite{Dig04}(dashed line). (right) Compilation of the 
$\sigma_{\rm J/\psi}/\sigma_{\rm DY}$ values measured in \mbox{p-A} and nucleus-nucleus
collisions at the SPS, rescaled, when necessary, to 158 GeV incident energy. The lines
indicate the results of a Glauber fit to the \mbox{p-A} data and the size of the error.
The full circle indicates the preliminary NA60 result for \mbox{p-A} collisions at 158
GeV.}
\label{fig:2}
\end{figure}

Several theoretical predictions for the \mbox{In-In} suppression pattern were formulated
before the NA60 experimental results became available. They include a model where the
anomalous suppression is due to interaction with hadronic comovers~\cite{Cap05}, another
where the effect of dissociation and regeneration in a fully thermalized QGP and in the
later hadronic stage is considered~\cite{Rap05}, and finally a model where parton
percolation occurs, with an onset at $N_{\rm part}\sim$140~\cite{Dig04}. It is
interesting to note that although these models were explicitely tuned on the already
available \mbox{Pb-Pb} results, none of them, as can be seen in Fig.~\ref{fig:2}(left) 
is able to quantitatively reproduce the \mbox{In-In} points (even if the overall size 
of the effect is reasonably reproduced). 
More recently, a study of the effect of a thermalized hadronic gas on the
\jpsi\ has been carried out~\cite{Mai05} in the frame of the Constituent Quark-Meson
model. The comparison of this calculation with data shows that for both \mbox{Pb-Pb} and
\mbox{In-In} hadronic effects alone cannot account for the observed anomalous
suppression.

\section{\jpsi\ {\bf suppression in \mbox{p-A} collisions}}
In order to claim that an anomalous suppression of the \jpsi\ has been observed, the
production in \mbox{p-A} collisions must be accurately known. Up to now, at SPS energy,
such knowledge came from measurements performed at 450 and 400 GeV, covering the rapidity
range $-0.5<y_{cm}<0.5$~\cite{Ale06}. Performing an analysis in the frame of the Glauber
model, two parameters are needed in order to fit the data, the cross section for
elementary collisions $(\sigma_{\rm J/\psi}/\sigma_{\rm DY})^{pp}_{E_{\rm 0}}$ at the energy
$E_{\rm 0}$ under consideration and the nuclear absorption cross section for the produced 
\jpsi, $\sigma^{\rm abs}_{\rm J/\psi}$. 
In order to obtain the expected yield for the rather different energy and
kinematical domain of the heavy-ion data (158 GeV/nucleon and 0$<y_{cm}<$1), a rescaling
of these parameters becomes necessary. Up to now, it was assumed that 
$\sigma^{\rm abs}_{\rm J/\psi}$ does not change as a function of the incident proton 
energy, and the cross section for elementary collision was rescaled using a procedure 
detailed in~\cite{Ale05,Bor05}. In order to avoid the systematic errors connected with 
this procedure, NA60 has measured \jpsi\ production in \mbox{p-A} collisions at 158 GeV,
in the same kinematical domain of the nucleus-nucleus data. A target box
containing nine subtargets, made of seven different materials 
(Be, Al, Cu, In, W, Pb and U) has been used, with 
the vertex spectrometer helping to recognize the target where the muon
pair has been produced. An analysis of the $A$-dependence of the \jpsi\ production cross
section would require of course a complete understanding of the local efficiency of the
vertex tracker, since its angular coverage for the various
targets is slightly different. Since this work is still in progress, for the moment a
preliminary analysis has been performed, using only the muon spectrometer information.
More in detail, one simply requires the extrapolation of the muon tracks to the target
region to lie inside the target box. This cut has not enough resolution to identify the
target where the interaction has taken place, but nevertheless efficiently rejects 
the background due to parasitic interactions outside the target region (e.g. in the 
hadron absorber). In this way, one can determine an average 
$\sigma_{\rm J/\psi}/\sigma_{\rm DY}$ ratio, that can be now plotted together with the
previous results, as a function of $L$, the mean thickness of nuclear matter crossed by
the produced \jpsi. Taking into account the nuclear composition of the target system, we
have, for this set of data, $\langle L\rangle = 3.4$ fm. In Fig.~\ref{fig:2}(right) we
show such a plot, where the closed circle indicates our preliminary result. It can be
seen that there is a very good agreement with the set of \mbox{p-A} data taken at higher
energy, and rescaled to 158 GeV. This result shows that the rescaling of the elementary
production cross section is indeed correct, and reinforces the claim that the 
suppression observed for \mbox{In-In} and \mbox{Pb-Pb} collisions is not compatible 
with a pure nuclear absorption scenario.

\section{\psip\ {\bf suppression in \mbox{p-A} and \mbox{In-In} collisions}}
\begin{figure}[h]
\centering
\resizebox{0.45\textwidth}{!}
{\includegraphics*{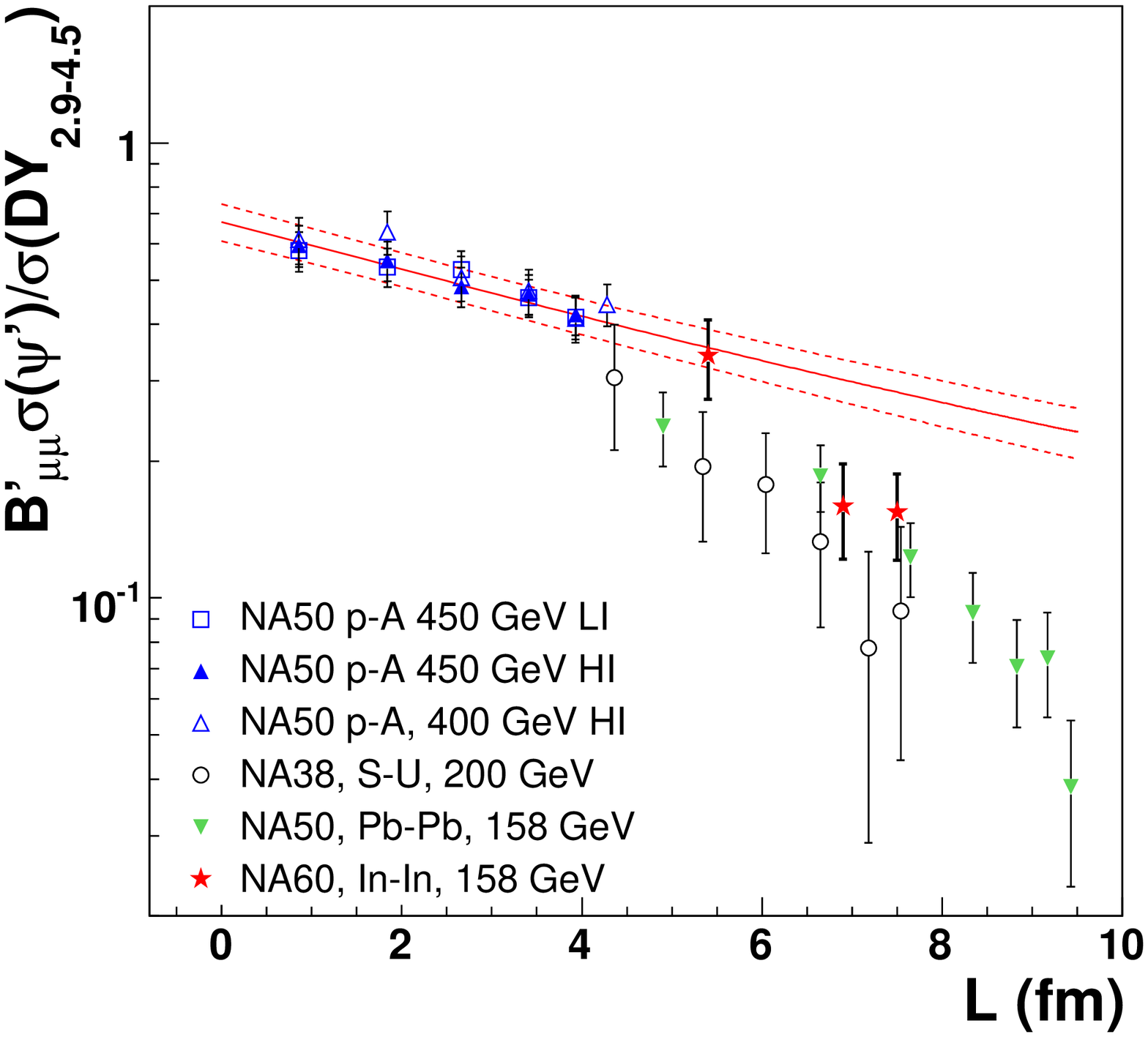}}
\resizebox{0.45\textwidth}{!}
{\includegraphics*{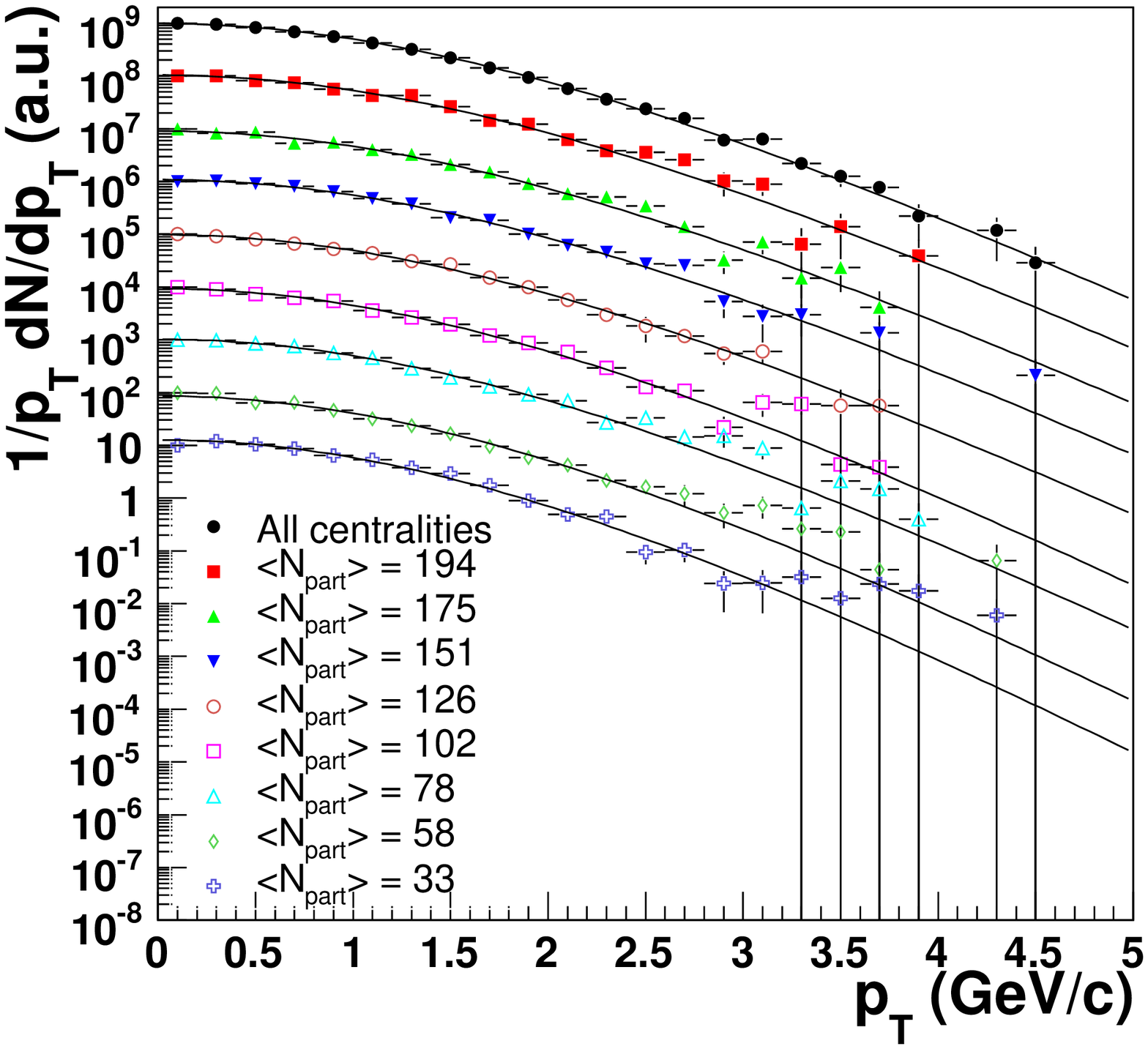}}
\caption{(left) Compilation of the 
$\sigma_{\psi '}/\sigma_{\rm DY}$ values measured in \mbox{p-A} and nucleus-nucleus
collisions at the SPS, rescaled, when necessary, to 158 GeV incident energy. The lines
indicate the results of a Glauber fit to the \mbox{p-A} data and the size of the error.
(right) The acceptance corrected $p_{\rm T}$ distributions for the \jpsi. The lines
correspond to fits to the function 
$1/p_{\rm T}{\rm d}N/{\rm d}p_{\rm T}=e^{-m_{\rm T}/T}$.} 
\label{fig:3}
\end{figure}

In addition to the \jpsi, NA60 can also detect the \psip\ through its muon pair decay.
The significance of this study, contrary to that of the \jpsi, is limited by the reduced 
available statistics (about 300 events in both \mbox{In-In} and \mbox{p-A}). In this paper we
present preliminary results on the ratio $\sigma_{\psi'}/\sigma_{\rm DY}$. For
\mbox{In-In} collisions, three centrality bins have been defined. The result is shown in
Fig.~\ref{fig:3}(left), as a function of $L$, and compared with previous findings by NA38
and NA50~\cite{Ram06}. One can again note, as for the \jpsi, that the production yield in \mbox{p-A}
collisions is fairly consistent with data taken at higher energies and then extrapolated
to 158 GeV. Concerning nucleus-nucleus data, the \mbox{In-In} points are found to be in 
fair agreement with the other measured nuclear systems. By comparing 
Fig.~\ref{fig:3}(left) with Fig.~\ref{fig:2}(right), we see that for the \psip\ the onset 
of an anomalous suppression might occur for more peripheral reactions than for the \jpsi.
Obviously, a larger statistics would be useful for reaching sharper conclusions.

\section{Kinematical distributions of the \jpsi\ produced {\bf in \mbox{In-In} collisions}}
\begin{figure}[h]
\centering
\resizebox{0.45\textwidth}{!}
{\includegraphics*{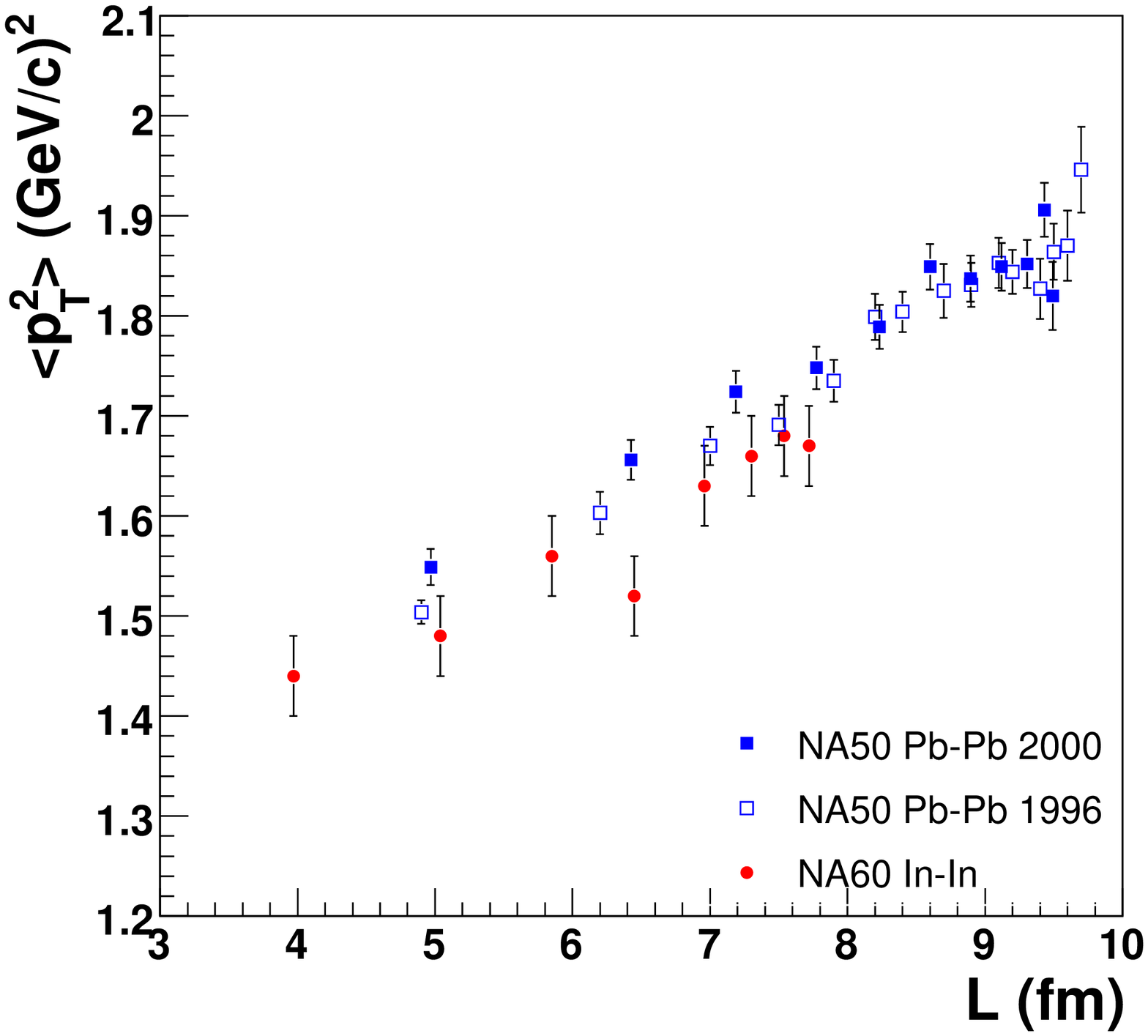}}
\resizebox{0.45\textwidth}{!}
{\includegraphics*{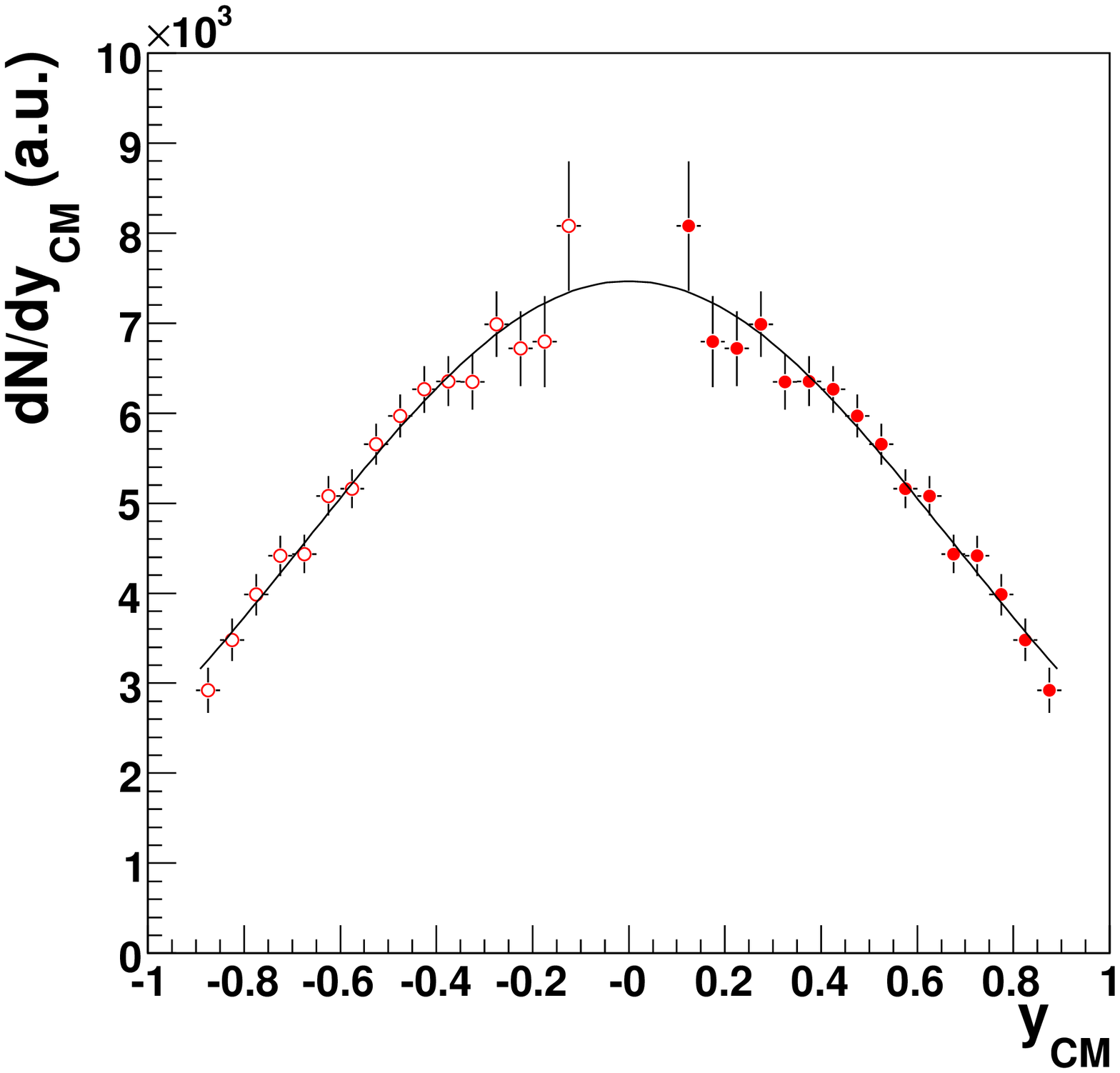}}
\caption{(left) The $L$-dependence of $\langle p_{\rm T}\rangle^2_{\rm J/\psi}$. Two sets of 
\mbox{Pb-Pb} data, corresponding to different NA50 data taking periods, are shown and 
compared with the NA60 result. (right) Centrality integrated $y$ distribution of the 
\jpsi\ produced in \mbox{In-In} collisions. The closed symbols represent the measured
points, that can be reflected around $y_{\rm CM}$=0 (open symbols).} 
\label{fig:4}
\end{figure}

Several important aspects of the charmonium production and interaction in nuclear
collisions can be addressed by studying the kinematical distributions of the produced
\jpsi. In particular, the transverse momentum distributions are expected
to be influenced by initial-state multiple scattering of the gluon~\cite{Bla89}, while the study of the rapidity 
distributions may be sensitive to details of the hadronization mechanism of the
$c\overline c$ pair~\cite{Vog00}. Furthermore, the study of the angular distribution of the 
decay products gives indications on the \jpsi\ polarization. This information is related
to the charmonium formation mechanism~\cite{Bra00} and may be affected by the presence of a deconfined
medium~\cite{Iof03}. From the experimental point of view, the \jpsi\ kinematical 
distributions have been
obtained by performing a 3-D acceptance correction. Events were generated with a flat
distribution in $p_{\rm T}, y, \cos\theta_{\rm H}$ ($\theta_{\rm H}$ is the decay angle of
the $\mu^+$, taken in the charmonium rest frame, with respect to the \jpsi\ direction in
the CM system), tracked and reconstructed in the set-up. A differential acceptance has
then been calculated in narrow bins (0.1 GeV/c in $p_{\rm T}$, 0.05 units in $y$ and 0.1
in $\cos\theta_{\rm H}$). Finally, the acceptance correction has been performed in the
kinematical domain where it is larger than 1\%. 
The whole procedure has been successfully tested on Monte-Carlo generated sample 
distributions; furthermore, the effect of the $\sim$3\% background below the \jpsi\ peak has been 
found to be negligible. In Fig.~\ref{fig:3}(right) we show the acceptance corrected 
$p_{\rm T}$ distributions, integrated over centrality and for various centrality bins. The
plots refer to the kinematical region 0.1$<y_{\rm CM}<$0.9, -0.4$<\cos\theta_{\rm H}<$0.4.
By fitting the distributions with the function 
$1/p_{\rm T}{\rm d}N/{\rm d}p_{\rm T}=e^{-m_{\rm T}/T}$ we obtain $T$ values increasing
with centrality and ranging from 204 to 234 MeV. For the centrality integrated
distribution we get $T$=231$\pm$2 MeV.
In Fig.~\ref{fig:4}(left) we show the increase of $\langle p_{\rm T}\rangle^2_{\rm J/\psi}$
with centrality. The \mbox{In-In} points are compared with \mbox{Pb-Pb} results obtained
by NA50~\cite{Top03}. Both data sets show a roughly linear increase of  
$\langle p_{T}\rangle^2_{\rm J/\psi}$ with $L$. Such a $p_{\rm T}$ broadening is
consistent with the occurrence of initial-state multiple scattering of the gluon. 
In Fig.~\ref{fig:4}(right) we show the centrality integrated $y$ distribution of the
produced \jpsi, where the points on the plot are obtained for 0$<p_{\rm T}<$5 GeV/c, 
-0.4$<\cos\theta_{\rm H}<$0.4. The distribution is well reproduced by a gaussian fit, with
$\sigma_{\rm y}$=0.68$\pm$0.02. Although not shown here, we find no significant dependence of the $y$ 
width on the centrality.

\begin{figure}[h]
\centering
\resizebox{0.30\textwidth}{!}
{\includegraphics*{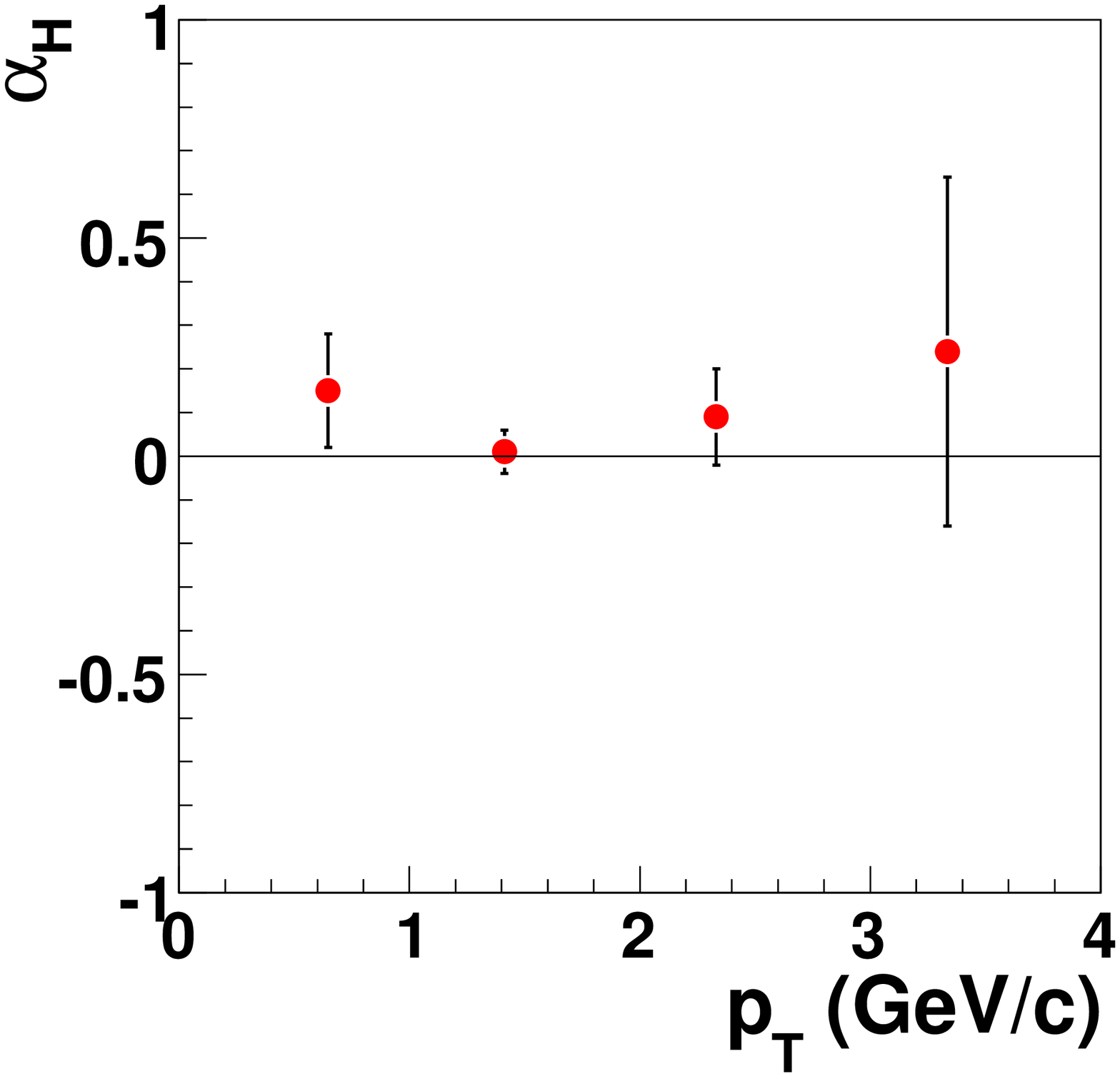}}
\resizebox{0.30\textwidth}{!}
{\includegraphics*{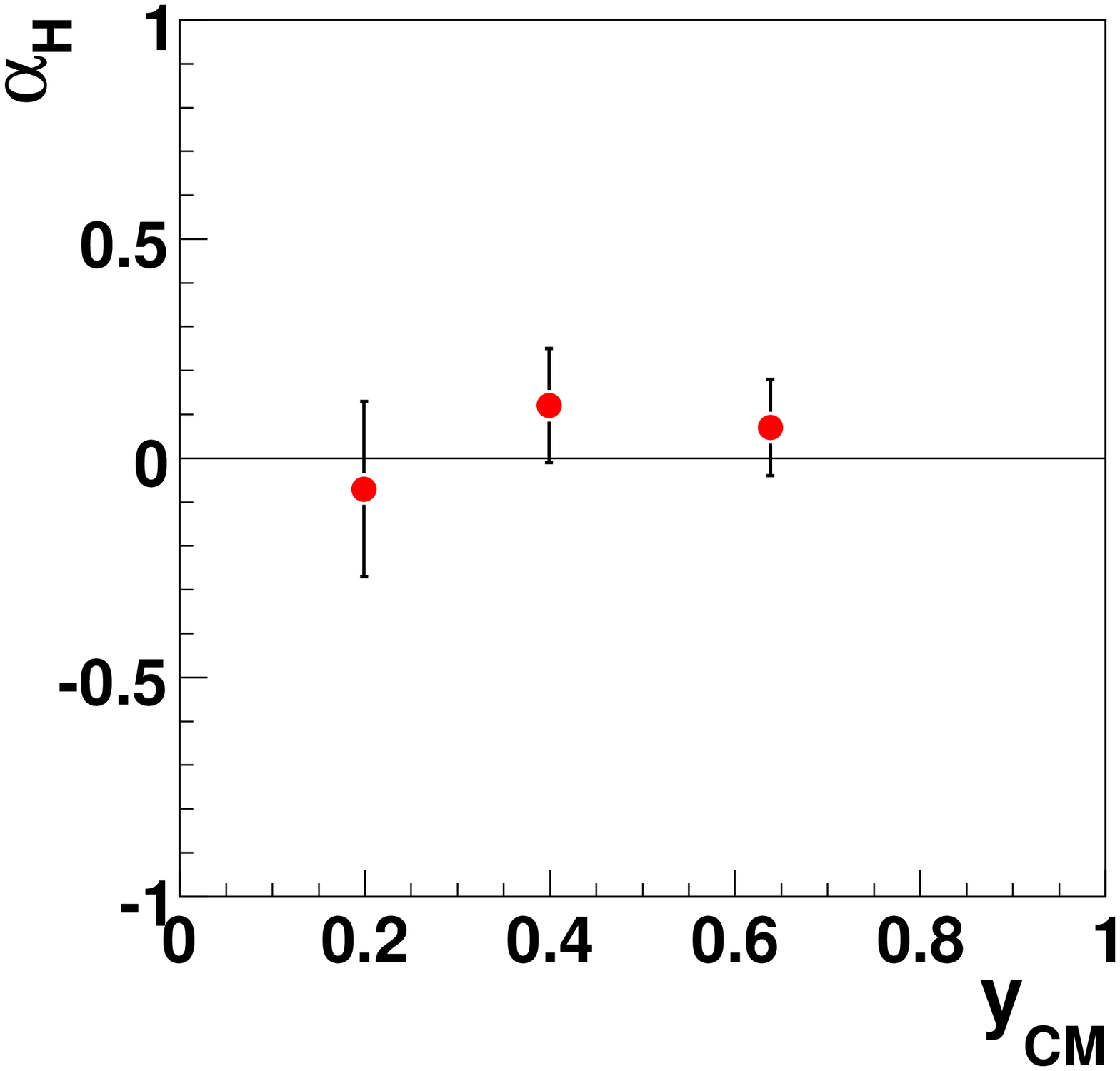}}
\resizebox{0.30\textwidth}{!}
{\includegraphics*{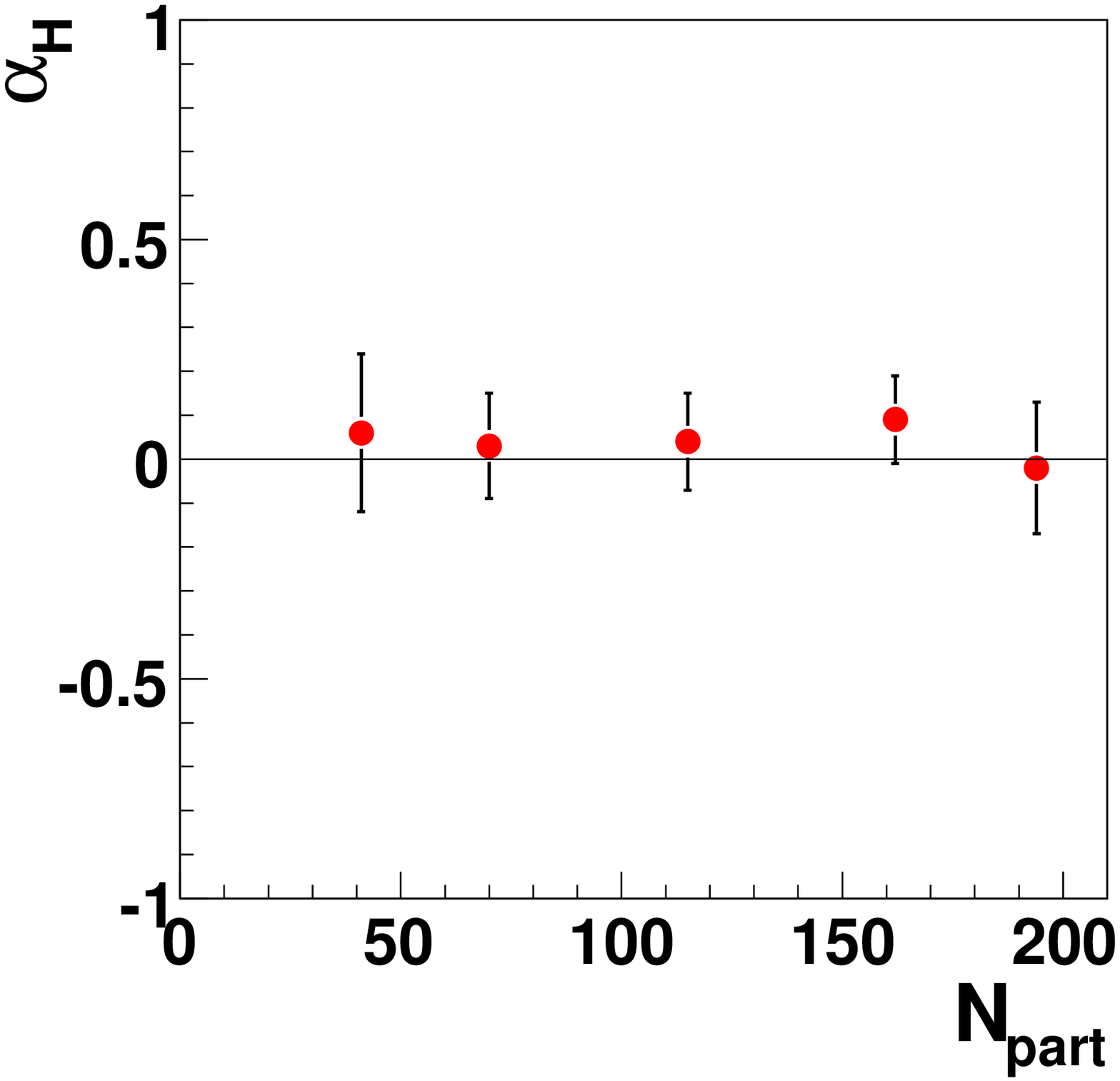}}
\caption{The \jpsi\ polarization, measured in \mbox{In-In}, as a function of $p_{\rm T}$
(left), $y$ (center) and $N_{\rm part}$ (right).} 
\label{fig:5}
\end{figure}

Finally, in Fig.~\ref{fig:5} we show the first results on the polarization of the \jpsi\ 
produced in heavy-ion collisions. The angular distributions of the decay products have been fitted
using the relation ${\rm d}\sigma/{\rm d}\cos\theta_{\rm H}=1+\alpha\cos^2\theta_{\rm H}$, where 
$\theta_{\rm H}$ is the polar decay angle of the positive muon in the helicity frame.
$\alpha$ values larger than 0 would indicate a transverse polarization for the \jpsi,
while $\alpha <$0 corresponds to a longitudinal polarization. NRQCD
calculations~\cite{Bra00} predict a
significant transverse polarization at high $p_{\rm T}$, that was not observed neither at fixed target
experiments~\cite{Cha03} nor at hadron
colliders~\cite{Aff00}. Furthermore, in nucleus-nucleus collisions, the occurrence of a transition to a
QGP might further enhance the observed polarization~\cite{Iof03}. Our preliminary results clearly 
indicate that, as a function of centrality, $p_{\rm T}$ and $y$, the \jpsi\ produced in 
\mbox{In-In} collisions do not exhibit any significant polarization.

\section{Conclusions}
The NA60 experiment has carried out a high-quality study of \jpsi\ production in
\mbox{In-In} collisions at the SPS. The results confirm, for a much lighter system, the
anomalous suppression seen in \mbox{Pb-Pb} collisions by NA50, with an onset at $N_{\rm
part}\sim$80, corresponding to $\epsilon_{\rm Bj}\sim$1.5 GeV/fm$^3$. First, preliminary 
results from \mbox{p-A} collisions at 158 GeV, the same energy of the nucleus-nucleus
data, strengthen our understanding of the effects of nuclear absorption on the
\jpsi\ and show that the results for peripheral \mbox{In-In} and \mbox{Pb-Pb} collisions
can be understood in terms of cold nuclear effects. Preliminary results on \psip\ also show 
an anomalous suppression, that sets in earlier than that of the \jpsi\ and increases 
for more central \mbox{In-In} collisions. Finally, the kinematical distributions
of the \jpsi\ produced in \mbox{In-In} have been studied. In particular, the first
preliminary results on the angular distributions of decay muons show that the \jpsi\ 
is produced without a significant polarization.

\end{document}